\documentclass[12pt, preprint]{aastex63}

\begin{document}

\title{Comparison of Enhanced Absorption in \ion{He}{1} 10830~\AA\ in Observations and Modeling During the Early Phase of a Solar Flare}

\author{Nengyi Huang}
\affiliation{Institute for Space Weather Sciences, New Jersey Institute of Technology, 323 Martin Luther King Blvd, Newark, NJ 07102-1982, USA}

\author{Viacheslav M. Sadykov}
\affiliation{NASA Ames Research Center, Moffett Field, CA 94035, USA}
\affiliation{Bay Area Environmental Research Institute, Moffett Field, CA 94035, USA}

\author{Yan Xu}
\affiliation{Institute for Space Weather Sciences,  New Jersey Institute of Technology, 323 Martin Luther King Blvd, Newark, NJ 07102-1982, USA}
\affiliation{Big Bear Solar Observatory, New Jersey Institute of Technology, 40386 North Shore Lane, Big Bear City, CA 92314-9672, USA}

\author{Ju Jing}
\affiliation{Institute for Space Weather Sciences,  New Jersey Institute of Technology, 323 Martin Luther King Blvd, Newark, NJ 07102-1982, USA}
\affiliation{Big Bear Solar Observatory, New Jersey Institute of Technology, 40386 North Shore Lane, Big Bear City, CA 92314-9672, USA}

\author{Haimin Wang}
\affiliation{Institute for Space Weather Sciences,  New Jersey Institute of Technology, 323 Martin Luther King Blvd, Newark, NJ 07102-1982, USA}
\affiliation{Big Bear Solar Observatory, New Jersey Institute of Technology, 40386 North Shore Lane, Big Bear City, CA 92314-9672, USA}

\submitjournal{ApJ L}
\accepted{10 Jun 2020}

\begin{abstract}
    The \ion{He}{1} 10830~\AA\ triplet is a very informative indicator of chromospheric activities as the helium is the second most abundant element in the solar atmosphere. Taking advantage of the high resolution of the 1.6 m Goode Solar Telescope (GST) at Big Bear Solar Observatory (BBSO), previous observations have shown clear evidence of the enhanced absorption, instead of typically-observed emission, for two M-class flares. In this study, we analyze the evolution of the \ion{He}{1} 10830~\AA\ emission in numerical models and compare it with observations. The models represent the RADYN simulation results obtained from the F-CHROMA database. We consider the models with the injected electron spectra parameters close to observational estimates for the 2013-August-17 flare event ($\delta{}=8$, $E_{c}=\{15,20\}\,keV$, $F=\{1\times{}10^{11},3\times{}10^{11}\}erg\cdot{}cm^{-2}$) in detail, as well as other available models. The modeling results agree well with observations, in the sense of both the maximum intensity decrease (-17.1\%, compared to the observed value of -13.7\%) and the trend of temporal variation (initial absorption phase followed by the emission). All models demonstrate the increased number densities and decreased ratio of the upper and lower level populations of \ion{He}{1} 10830~\AA\ transition in the initial phase, which enhances the opacity and forms an absorption feature. Models suggest that the temperatures and free electron densities at heights of 1.3-1.5\,Mm should be larger than $\sim{}$10$^4$\,K and 6$\times{}$10$^{11}$\,cm$^{-3}$ thresholds for the line to start being in emission.
\end{abstract}

\section{Introduction}
The \ion{He}{1} 10830~\AA\ triplet, centered at wavelengths of 10829.081~\AA, 10830.250~\AA\, and 10830.341~\AA\ respectively is relatively weak in comparison with other prominent chromospheric lines, but represents a powerful diagnostic tool for the chromospheric processes. Since the formation requirements of this line match the condition of upper chromosphere and lower corona, it provides rich information of many observational phenomena, including flares, coronal mass ejections (CMEs), solar oscillations, magnetic field dynamics, and filaments/prominences \citep{Zirin1966, Harvey1971, Landman1976, Harvey1984, You1992, Fleck1994, Rueedi1995, Lin1998, Xu2016, Anan2018, Libbrecht2019}.

Usually, the \ion{He}{1} 10830~\AA\ line is seen in absorption against the bright solar disk: in filaments, H$\alpha$ network, and coronal holes \citep{Zirin1966, Tandberg-Hanssen1962, Harvey1984, Harvey2002}. This line is believed to form in a wide range of heights, from 1.1 Mm \citep{Muglach2001} to 2.4 Mm \citep{Schmidt1994}, and corresponds to the transition between 2$^{3}$S$_{1}$ and 2$^{3}$P$_{0,1,2}$ atomic levels of non-ionized helium. To populate electrons in helium atoms from the ground state to those higher triplet states, high temperature and density are required \citep{Mohler1956,Zirin1966, Zirin1975}. Such conditions can be provided by radiative or collisional mechanisms \citep{Athay1960}. During flares, the \ion{He}{1} 10830~\AA\ line turns into emission as the majority of other spectral lines do. Strong emission, several times larger than the background intensity, and corresponding enhanced line broadening, have been reported in several flares of the GOES classes ranging from C-class to X20 \citep{You1992, Penn1995, Penn2006, Zeng2014}. 

In contrast to the typically-observed enhanced emission, there are reports of the enhanced absorption of spectral lines or continua, also known as ``negative flares'' \citep{Flesch1974,Henoux1990,Zirin1980}. In particular, for the \ion{He}{1} 10830~\AA\ line, \citet{Xu2016} presented the analysis of two M-class flares showing enhanced absorption appearing on the leading edge of the flare ribbons. Since the two ribbons propagate away from the local magnetic polarity inversion line, their leading edges represent the footpoints of the newly-reconnected magnetic loops. In other words, the enhanced absorption occurs at the very beginning of the localized flare heating process.

Theoretical studies mention three line formation mechanisms, namely collisional-activation mechanism, photo-ionization recombination mechanism, and collisional-ionization recombination mechanism \citep[CM, PRM, and CRM correspondingly,][]{Goldberg1939,Andretta1997,Centeno2009}. Under flare conditions, excessive energy input leads to the enhanced absorption of the line at the initial phase, and the strong emission afterward. Such behavior has been studied numerically by \citet{Ding2005}, by assuming non-LTE statistical equilibrium approximation for the atomic level populations and hydrostatic equilibrium of the atmosphere in calculations. In this study, we use the advantage of the state-of-the-art RADYN \citep{Carlsson1997,Allred2005,Cheng2013,Allred2015} radiative hydrodynamics code results publicly-available under F-CHROMA project. The RADYN models the dynamically-evolving atmospheric response to the energy deposit as a function of time, under the non-LTE non-equilibrium condition (NEC). We make a comparison between one of the observed \ion{He}{1} negative flares by \citet{Xu2016} with the \ion{He}{1} 10830~\AA\ line emission from RADYN simulations with the closest-matching electron beam heating parameters. Then we discuss the possible physical conditions in the chromosphere in reaction to electron beam heating that generate the enhanced \ion{He}{1} 10830~\AA\ absorption.

\section{Description of Observations and Models}

\citet{Xu2016} presented two M-class flares with the negative contrast in \ion{He}{1} 10830~\AA, observed by the 1.6 m GST at BBSO. The high-resolution observations show enhanced absorption in a very narrow spatial region (about 500 km), in front of the normal flare ribbon with strong emission. The maximum decrease in intensity is about -13.7\%, comparing with the pre-flare condition. The duration of the intensity drop is about 90 s. 

It is well accepted that the energetic electrons precipitating into the atmosphere along the magnetic field lines represent one of the mechanisms of the lower atmosphere heating during flares. To understand the details of how the atmosphere is heated, it is helpful to consider radiative hydrodynamic modeling. RADYN code is one such modeling approach and has been widely used in the community. By assuming non-LTE NEC, RADYN solves hydrodynamic equations and the radiative transfer of the dominating atoms in the solar atmosphere, including helium. Thanks to the F-CHROMA project, a database of RADYN simulations of flares driven by different electron beams is publicly available online\footnote{http://www.fchroma.org/}. In these models, the atmosphere heating is caused by the precipitating electron beam with the power-law spectra described by the power-law index ($\delta$), total energy flux (F) and lower energy cut-off (E$_{c}$). The output consists of the intensities in different spectral windows, including both spectral lines and continua, the corresponding energy terms, and the stratification of physical parameters of the atmosphere. Each F-CHROMA model has 500 time steps with 0.1\,s time interval. The electron heating lasts for 20\,s (200 time steps) and follows a triangular shape, in which the electron flux increases monotonically from zero to the peak in 100 steps (for instance, the electron energy flux, F, reaches $1 \times 10^{10}erg\cdot cm^{-2}\cdot{}s^{-1}$ value for the model with the total deposited energy of E$_{tot}$ = $1 \times 10^{11}erg\cdot cm^{-2}$) and then decreases back to zero in the next 100 time steps. The starting atmosphere is fixed to VAL-C~\citep{Vernazza1981}, and the energetic electron transport is solved using the Fokker-Planck equation. In RADYN calculation, the lowest 5 energy levels of \ion{He}{1}, lowest 3 energy levels of \ion{He}{2} and the continuum helium are included. These include the ground state helium, the orthohelium states which generate the \ion{He}{1} 10830~\AA\ line, and the excited helium.

In order to achieve transitions between 2$^{3}$S$_{1}$ and 2$^{3}$P$_{0,1,2}$ levels, the helium atoms need to be populated from parahelium (with two electrons spinning in the opposite direction) ground state to the corresponding triplet states of orthohelium (with two electrons spinning in the same direction). According to Pauli's Rule, the direct activating mechanism is limited to CM, which enables the change of the total spin number. On the other hand, recombination following ionization is also possible to produce orthohelium to generate triplets. Either the PRM by extreme ultraviolet (EUV) backwarming effect from heated corona or CRM by high energy non-thermal electron beams should be considered. RADYN is comprehensive for the simulation of \ion{He}{1} 10830~{\AA} since it not only includes the transitions that generate the \ion{He}{1} 10830 \AA\ triplet explicitly from the numerical perspective, but also considers effects important for the line formation to a certain extent. The F-CHROMA RADYN runs utilized in this work consider the photon-ionization from coronal heating (EUV radiative backwarming) to enable the PRM. 
The non-thermal as well as thermal collisional ionization rates for the \ion{He}{1} and \ion{He}{2} species contributing to the CRM are included explicitly in the models~\citep{Allred2015}, but only thermal collisional excitation rates of \ion{He}{1} are included in CM.

The electron beam heating parameters for RADYN can be estimated from hard X-ray (HXR) observations taken by the Reuven Ramaty High Energy Solar Spectroscopic Imager \citep[RHESSI,][]{RHESSI}. One of the two flares studied by \citet{Xu2016} was partially covered by RHESSI on 2013-Aug-17. The spectrum of the flare in the initial phase, i.e., during 18:33:16 UT - 18:33:20 UT, was used to retrieve the quantities to describe the precipitating electron beam. The spectrum was fitted using the combination of thermal and non-thermal thick-target (version 2) models in OSPEX. The parameters of the non-thermal distribution of electrons are found to be $\delta$ = 8.23, E$_{c}$ = 16.9 keV, and total number of electrons = 6.58$E^{35}$ electrons s$^{-1}$. Considering the electron precipitation area of $\sim{}10^{18}~cm^{2}$ (estimated as the area of RHESSI 25-50\,keV sources reconstructed using CLEAN algorithm), the peak energy flux, F, is about $10^{10} erg\cdot cm^{-2}\cdot s^{-1}$. Since these parameters are derived using the HXR spectrum not obtained simultaneously with the enhanced absorption of \ion{He}{1} 10830 \AA, we consider the following multiple values of the parameters of the F-CHROMA model grid: $\delta{}=8$,  E$_{c}$ = 15 keV and 20 keV, E$_{tot}$ = $1 \times 10^{11}erg\cdot cm^{-2}$ and $3 \times 10^{11}erg\cdot cm^{-2}$.

\section{Results}

The F-CHROMA database includes 80 sets of RADYN runs, with different characteristics of electron beams as inputs. As mentioned previously, the exact match of parameters from the RHESSI observation is not available in the F-CHROMA database. Because of that we choose to investigate multiple sets of RADYN runs. RADYN results from four sets of beam parameters with $\delta$ =  8, E$_{c}$ =  15 keV and 20 keV, and E$_{tot}$ = $1 \times 10^{11}$ erg $cm^{-2}$ and $3 \times 10^{11}$ erg $cm^{-2}$ (also listed in Table~\ref{model}) were considered in detail. We also use the model ``val3c\_d8\_1.0e11\_t20s\_20keV'' (with $\delta$ = 8, E$_{tot}$ = $1 \times 10^{11}erg\cdot cm^{-2}$ and E$_{c}$ = 20 keV) as a representative model which has the closest values to the parameters derived from RHESSI observations in terms of the deposited electron spectra.

\begin{figure}
    \centering
    \includegraphics[scale=.79]{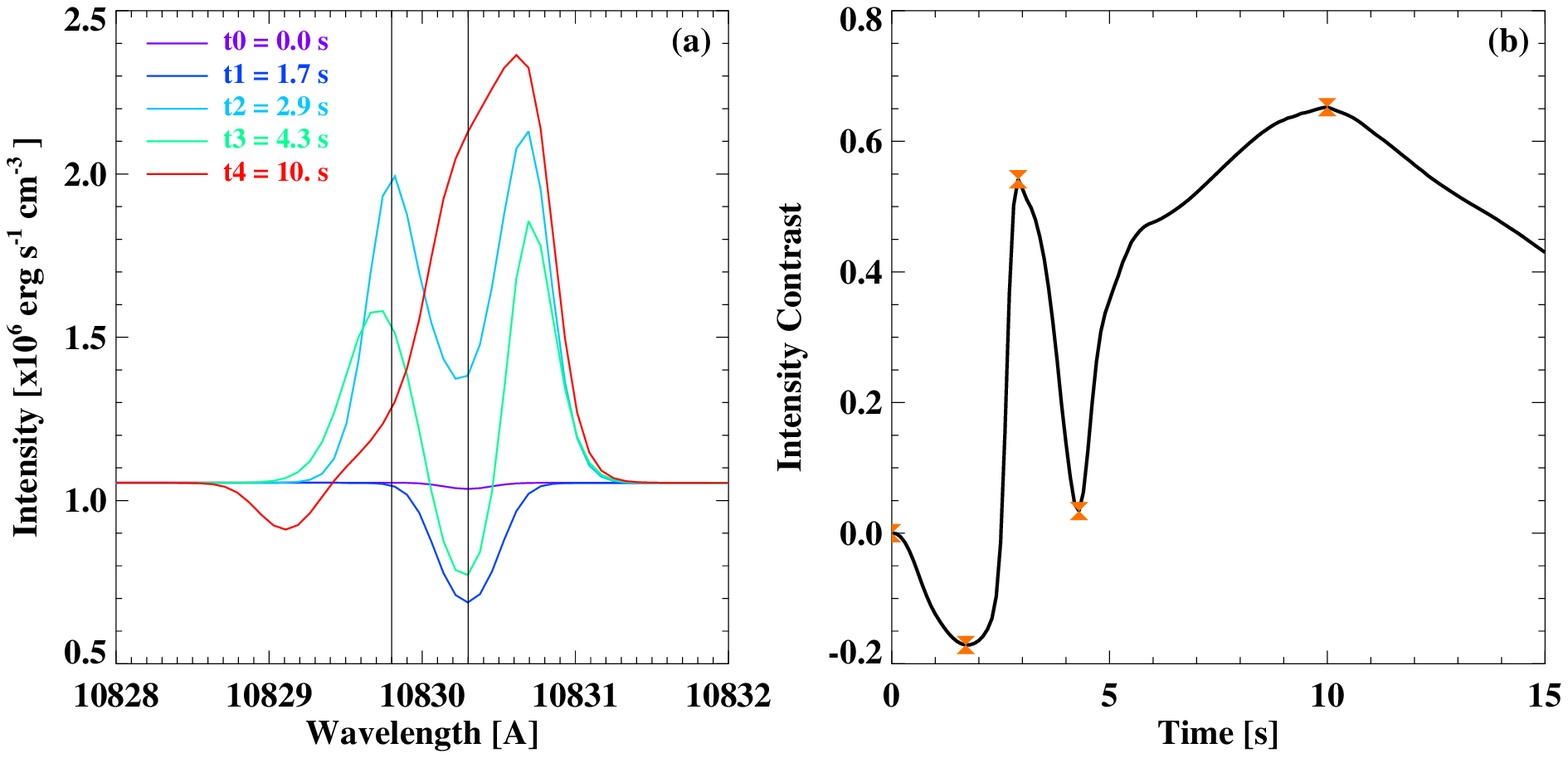}
    \includegraphics[width=0.9\columnwidth]{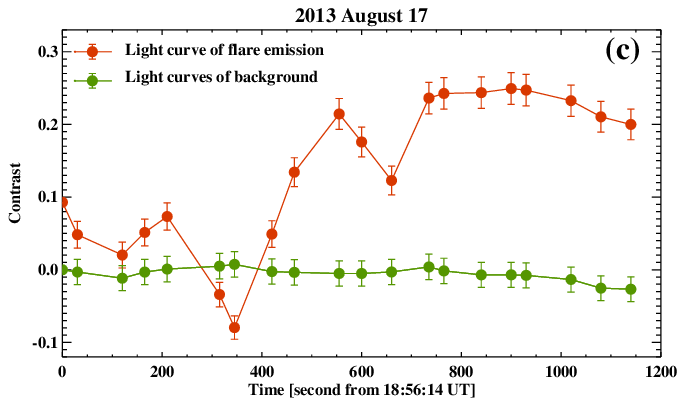}
    \caption{Panel (a): \ion{He}{1} 10830~\AA\ line profiles at different times, for the RADYN model with $\delta$ = 8, total energy of $1 \times 10^{11} erg ~ cm^{-2}$ and low energy cutoff of 20 kev. Enhanced absorption is seen at t = 1.7\,s and turn into strong emission at t =2.9\,s. Panel (b): The modeled contrast light curve obtained within the same spectral window as for the observation, shown between the black vertical lines in the left panel. Panel (c): Reproduced light curves of the flaring area and a quiet Sun area (background) from BBSO/GST observation following \citet{Xu2016}.}
    \label{line_n_timeprofile}
\end{figure}

Examples of \ion{He}{1} 10830~{\AA} line profiles at five different times for this model are shown in the panel (a) of Figure~\ref{line_n_timeprofile}. The BBSO/GST observations were obtained using a tunable Lyot Filter \citep{Cao2010} at a fixed bandpass at the blue wing of \ion{He}{1} line at 10830.05 $\pm$ 0.25~\AA. To make a comparison between modeling and observation, the same spectral window is used for modeled spectra, as shown by the two vertical lines in panel (a) of Figure~\ref{line_n_timeprofile}. By integrating the intensities within this bandpass at different times, the modeled light curve is plotted in panel (b). It is normalized with respect to the first point, which is considered as the pre-flare condition. In Figure 4 of \citet{Xu2016}, the light curve obtained from BBSO/GST observation shows that the enhanced absorption occurred at the initial stage of the flare, followed by emission afterward. We also reproduce this figure in the panel (c) of the Figure~\ref{line_n_timeprofile}. It is obvious that the modeled results show a rapid drop of intensity followed by emission, which is similar to the observed temporal variation pattern, although their timescales are different. More importantly, the maximum dimming of the modeled intensity is about -17.1$\%$ in contrast to the pre-flare level, which is comparable to the value of -13.7$\%$ found in the observation. In addition to the initial absorption feature, we can also see the second dip of the passband emission, which also agrees with the observed behavior. On the other hand, the timescale of the modeled intensity differs from the observations. The duration of the enhanced absorption lasts about 90 s and the modeled absorption vanishes in several seconds. This discrepancy is likely a result of short timescales of the electron heating (20\,s) in the F-CHROMA RADYN runs. The previous study suggested that the timescale of an electron thread heating the atmosphere is on the order of 200 s, and a shorter time span can lead to overrapid evolution \citep{Warren2006}.

\begin{figure}
    \centering
    \includegraphics[scale=.8]{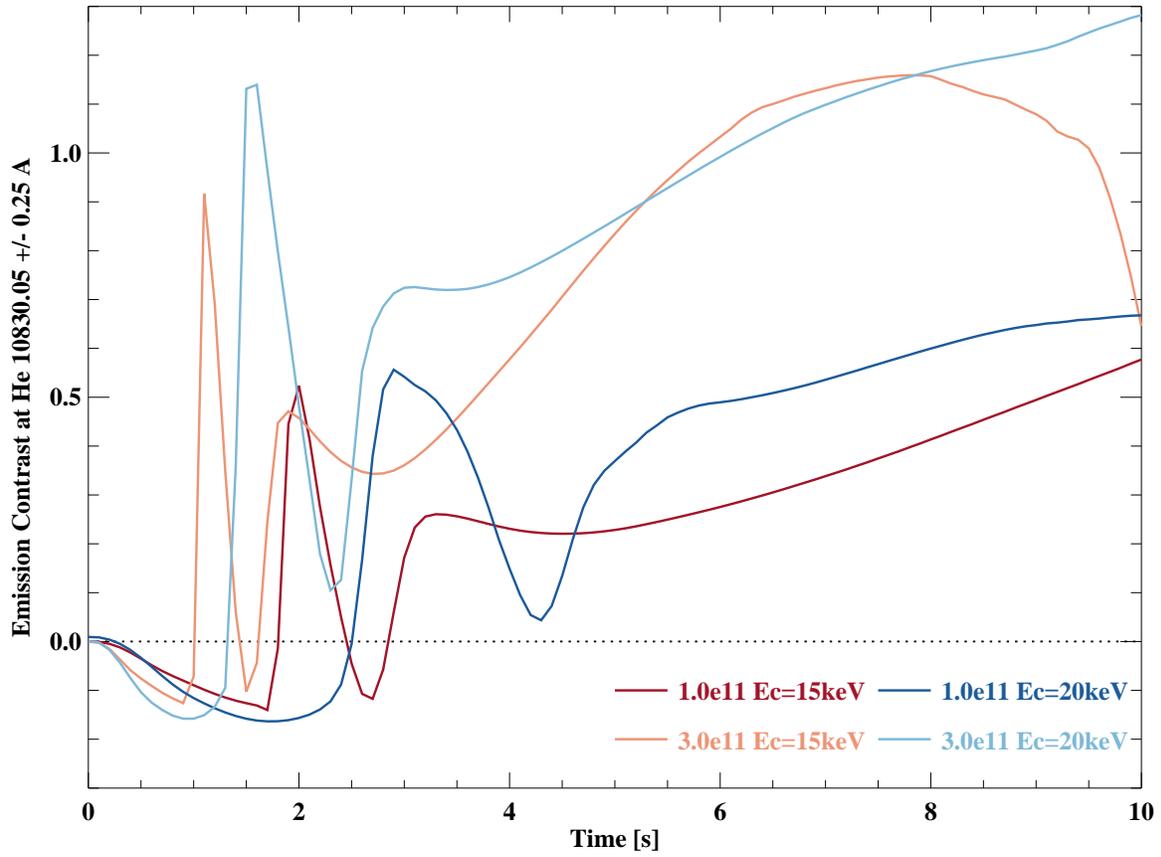}
    \caption{The normalized light curves obtained for 4 F-CHROMA models closest to the observations in terms of the HXR spectra parameters. The intensity was integrated over 10830.05 $\pm$ 0.25~\AA\ spectral window and normalized with respect to the a same reference level of $1.037 \times 10^{6} erg ~ cm^{-2} s^{-1}$.}
    \label{4_models}
\end{figure}

Figure~\ref{4_models} shows the light curves of \ion{He}{1} 10830~\AA\ blue wing for the four models. They are normalized to the first points (which all have the same intensity of $1.037 \times 10^{6} erg ~ cm^{-2} s^{-1}$). As we can see, all four models give an enhanced absorption right after the start of the injection of electron beams and turn to emission as most of the solar flares do due to continued precipitation of electrons later in the heating. On the other hand, the duration of enhanced absorption differs from model to model. The trend is that the lower F tends to have a longer duration. Moreover, they are accompanied by weaker emission afterward. The E$_{c}$ seems to affect the second dip--the short-term decrease of emission. The lower E$_{c}$ is, the stronger the second dip will be. 

\begin{table}
\begin{center}
\caption{Parameters of the injected electron spectra from the F-CHROMA RADYN grid selected according to the observational HXR spectrum.}
\label{model}

\begin{tabular}{lccc}
\\
 \hline\noalign{\smallskip}
F-CHROMA Model Number       &	Total Energy ($erg ~ cm^{-2} $)		&	Low-energy Cutoff (keV)	&        Power-law Index	\\
 \hline\noalign{\smallskip}

24		&		$1 \times 10^{11}$		&		15		&		8		\\
42		&		$1 \times 10^{11}$		&		20		&		8		\\
30		&		$3 \times 10^{11}$		&		15		&		8		\\
48		&		$3 \times 10^{11}$		&		20		&		8		\\

 \hline\noalign{\smallskip}\hline
\end{tabular}
\end{center}
\end{table}

To understand the conditions of the atmosphere corresponding to the line absorption and emission phases, we illustrate the temperatures, electron densities, population ratios for levels forming the \ion{He}{1} 10830~\AA\ transition, and the contribution function averaged in 10830.05$\pm$0.25\,\AA\ passband, for the previously discussed model ``val3c\_d8\_1.0e11\_t20s\_20keV'' in Figure~\ref{snapshot}. As one can see, both the temperature and the density of free electrons become enhanced even during the initial absorption phase of the line evolution. Interestingly, although both number densities of He 2$^{3}$S$_{1}$ and 2$^{3}$P$_{0,1,2}$ levels significantly increase, the ratio of populations of \ion{He}{1} 10830~\AA\ upper level to lower level (n$_{upper}$/n$_{lower}$) decreases at the heights above $\sim$1.35\,Mm during the line absorption phase, and significantly increases when the line is in emission. The contribution function presented in Figure~\ref{snapshot} also has a significant component at heights above 1.0\,Mm during the line absorption/emission phase.

\section{Discussion}

In this study, we presented the analysis of numerical models of \ion{He}{1}~10830~\AA\ line emission during the flare heating and compare them with BBSO/GST observations of the M-class flare. We found that: 1) An enhanced absorption is reproduced by RADYN simulation at the initial stage of the flare; 2) The level of modeled absorption is about 17\%, which is comparable with the observed level of 13\%; 3) A second dip, which was neglected by the previous models but noticed by observations, is also reproduced by the considered models and motivates further analysis.

\begin{figure}
    \centering
    F-CHROMA model ``val3c\_d8\_1.0e11\_t20s\_20kev\_fp'' \\
    \includegraphics[width=1.0\linewidth]{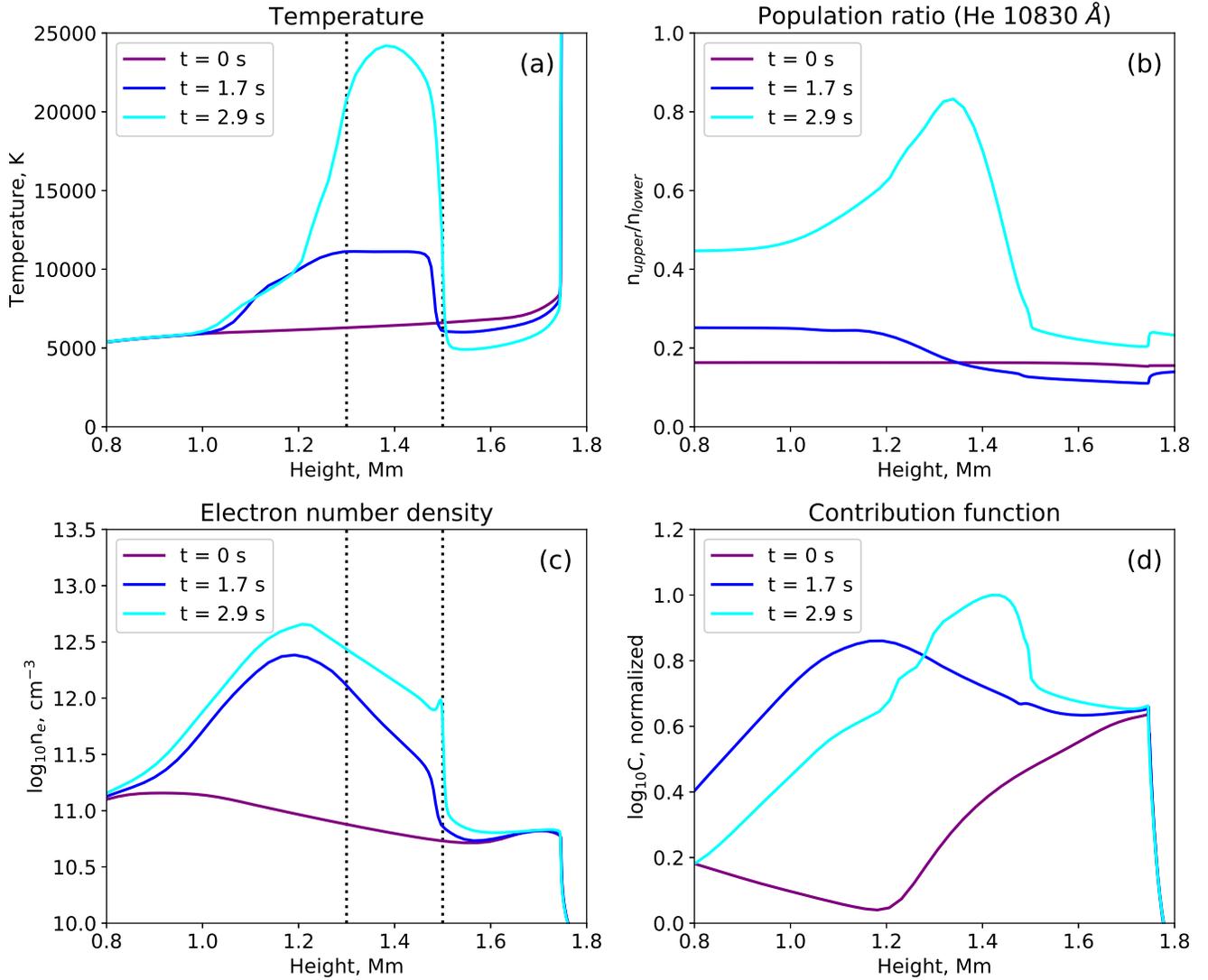}
    \caption{Illustration of the (a) temperature profiles, (b) population ratios for He levels forming \ion{He}{1} 10830 \AA\ transition, (c) electron number density profiles, and (d) normalized \ion{He}{1} 10830~\AA\ line contribution functions averaged over the 10830.05$\pm$0.25\,\AA\ passband for the selected RADYN model. The electron beam parameters in the model are $\delta$ = 8, $E_{tot}$ = $1 \times 10^{11} erg ~ cm^{-2}$ and $E_{c}$ = 20 kev. Plots are colored according to the timings using the same color code as in Figure~\ref{line_n_timeprofile}a: the beginning time (black), the time of deepest absorption t = 1.7\,s ( dark cyan) and the time that the line turns into strong emission t= 2.9\,s (cyan). Dotted} vertical lines in panels (a) and (c) mark the 1.3-1.5\,Mm height range.
    \label{snapshot}
\end{figure}

According to \citet{Zirin1988}, the \ion{He}{1} D3 line turns from absorption to emission at high temperature (T$>2 \times 10^{4}$ K) and plasma density (N$>5 \times 10^{12}$ cm$^{-3}$). In principle, the populations, at the two metastable states of 2$^{3}$P$_{0,1,2}$ and 3$^{3}$D$_{1,2,3}$, determine whether the D3 line is absorption or emission. A straightforward hypothesis is that similar thresholds may exist for the \ion{He}{1} 10830~\AA\ line. The outputs of a RADYN run include the condition of the heated atmosphere (i.e. `snapshot') and the corresponding spectral line/continuum profiles emitted from such atmosphere snapshots. Figure~\ref{snapshot} shows an example of the atmospheric stratification in terms of the temperature (a), the ratio of the atomic level populations forming \ion{He}{1} 10830~\AA\ transition (b), and electron densities (c). The colors indicate the timing, with purple to dark cyan colors representing the system time from 0 s to 1.7 s. According to the literature, the formation heights of \ion{He}{1} 10830~\AA\ were found to range from 1 Mm to 1.5 Mm \citep{Muglach2001, Schmidt1994}. As we see, this agrees well with the behavior of the passband contribution function presented in the panel (d) of the Figure~\ref{snapshot}, which becomes significantly enhanced above 1\,Mm, with the clear peak at $\sim$1.4\,Mm during the emission phase. In the first couple of seconds of the flare, the electron density within the same height range increases rapidly from about $1.4 \times 10^{11} cm^{-3}$ to $2.4 \times 10^{12} cm^{-3}$, while the temperature changes more constantly from about 6,500 K to 11,000 K. During the following seconds (dark cyan to cyan, t~=~1.7~s to t~=~2.9~s) the number density increases less intensively to about $4.4 \times 10^{12} cm^{-3}$, while the temperature keeps increasing constantly to around 23,000 K. During this period, the integrated blue wing of \ion{He}{1} 10830~{\AA} changes from enhanced absorption to emission. Therefore, the inferred watershed of emission and absorption in the \ion{He}{1} 10830~\AA\ line for the representative run is the condition of T$\gtrapprox{}2 \times 10^{4}$ K and n$_{e}$$\gtrapprox{}4 \times 10^{12} cm^{-3}$.

\begin{figure}
    \centering
    \includegraphics[width=1.0\linewidth]{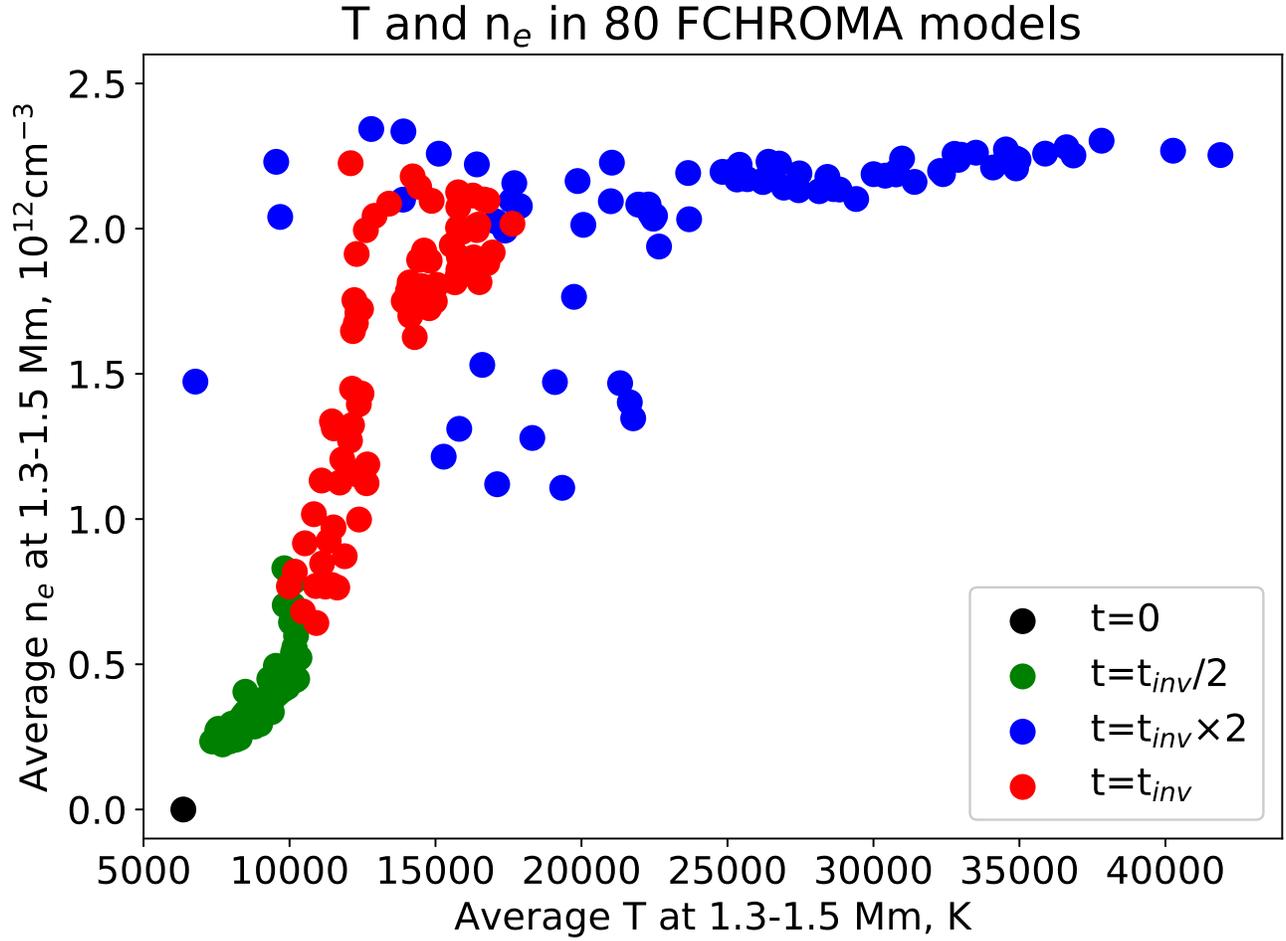}
    \caption{Distribution of temperatures (T) and electron number densities (n$_e$) averaged at 1.3-1.5\,Mm height for 80 available RADYN F-CHROMA models. Red points correspond to the T and n$_{e}$ values at the time when the line intensity obtained at 10830.05$\pm$0.25\,\AA\ turns from absorption to emission ($t_{inv}$), green points~--- to twice shorter time (t$_{inv}$/2), blue points~--- to twice longer time (t$_{inv}\times$2). Black point marks the initial atmospheric conditions for each run.}
    \label{TNEscatterplot}
\end{figure}

To confirm the existence of thresholds in general, we consider all 80 RADYN models available in F-CHROMA. It is necessary to mention that all models demonstrate the initial absorption in the \ion{He}{1} 10830~\AA\ passband, followed by the emission. We consider the temperatures and electron densities averaged at 1.3-1.5\,Mm heights for these models at the time when the absorption changes to the emission (t$_{inv}$), as well as at the twice shorter and longer times. The scatter plot presented in Figure~\ref{TNEscatterplot} demonstrates that temperatures and electron densities during t$_{inv}$ are distinguishable from those during the absorption and emission phases. On average, the temperature at 1.3-1.5\,Mm heights should be above 1.3$\times{}10^{4}$\,K, and the electron density should be above 1.4$\times{}10^{12}$\,$cm^{-3}$ for the line to turn into emission. It is also important to mention that the temperatures and electron densities averaged over other heights demonstrate less clearer separation between the absorption and emission phases with respect to 1.3-1.5\,Mm range.

A high ionization-recombination rate would overpopulate the metastable state of orthohelium, 2$^{3}$S$_{1}$, which is the lower level of \ion{He}{1} 10830~{\AA} transition (see Figure~\ref{snapshot}c for details), and then enhance the absorption. Previous studies often focus on CRM \citep{Ding2005, Xu2016} for flare emission. Our study confirms that the non-thermal atomic level populations corresponding to the \ion{He}{1} 10830~{\AA} transition increase fast during the absorption enhancement at the formation heights of \ion{He}{1} 10830. This would increase the collisional-ionization and recombination rate and overpopulate the lower level of the \ion{He}{1} 10830~{\AA} transition with respect to pre-flare (initial) conditions, as evident in Figure~\ref{snapshot}b. On the other hand, PRM were believed to be dominant in the formation of \ion{He}{1} 10830~\AA\ line. During the flare initial phase, backwarming effect, resulting in stronger photon-ionization effect, would also contribute to the overpopulation of helium 2$^{3}$S$_{1}$ state. When the upper chromosphere was heated, the higher rate of direct collisional excitation by thermal electrons would raise the occupation of both excited energy level and turn the line into emission. The temperature we retrieved from the model, 23000 K, agreed with the theoretical required temperature of Lyman plateau around 25,000 K \citep{Milkey1973}.

In this study, the electron precipitation area is estimated as a RHESSI 25-50\,keV HXR source area (within 50\% contour level) reconstructed with CLEAN algorithm. 
This is a widely-used but simplified approach that likely leads to overestimated precipitation areas. Correspondingly, the derived energy flux of F = $10^{10}~erg\cdot cm^{-2}\cdot s^{-1}$ is likely a lower limit. The precipitating electrons are confined by the magnetic field lines, which are converging from the corona to the chromosphere. As a consequence, the flaring areas become smaller in the deeper atmosphere \citep{Xu2012a}. For instance, considering the width of the flare ribbon measured by \citet{Xu2016} and the ribbon length observed by SDO/AIA 1700~\AA, the source area is about $3 \times 10^{17}~cm^{2}$. Consequently, the energy flux becomes F = $\sim10^{11}~erg\cdot cm^{-2}\cdot s^{-1}$ as estimated from observations. This should be compared against the F-CHROMA models of at least $E_{tot}$ = $1 \times 10^{12}~erg ~ cm^{-2}$, $\delta{}=8$,  E$_{c}$ = 15 keV and 20 keV, which are not currently available in the database. On the other hand, the total energy flux does not appear to impact the presence of the absorption or the formation condition of the He\,I\,10830\,{\AA} line. The higher energy flux may be able to bring the absorption forward due to its faster electron injection rate. For a better understanding of the impact of different electron injection on the time evolution of \ion{He}{1} 10830 line, a further study focusing on evolution and expanded to more models is required.

\begin{acknowledgements}\par
We thank Dr. Mats Carlsson for helping with interpretation of the the F-CHROMA models, and Drs. Graham S. Kerr and Vanessa Polito for helpful discussions. We would like to thank the anonymous referee for the very valuable comments in improving this work. This work is supported by NSF under grants AGS 1539791, 1821294 and 1954737, and by the NASA under grants 80NSSC19K0257 and 80NSSC19K0859. VMS is supported by the NASA grants NNX12AD05A and NNX16AP05H. F-CHROMA project is funded by the European Community's Seventh Framework Programme (FP7/2007-2013) under grant agreement no. 606862 (F-CHROMA). 

\end{acknowledgements}

\end{document}